\pdfoutput=1
\documentclass[manuscript]{acmart}
\AtBeginDocument{%
  }

\copyrightyear{2023} 
\acmYear{2023} 
\setcopyright{acmlicensed}\acmConference[RecSys '23]{Seventeenth ACM Conference on Recommender Systems}{September 18--22, 2023}{Singapore, Singapore}
\acmBooktitle{Seventeenth ACM Conference on Recommender Systems (RecSys '23), September 18--22, 2023, Singapore, Singapore}
\acmPrice{15.00}
\acmDOI{10.1145/3604915.3608821}
\acmISBN{979-8-4007-0241-9/23/09}

\usepackage{subfigure}
\usepackage{multirow}
\begin{document}
\title[Statistics-Driven Pre-training for Stabilizing Sequential Recommendation Model]{Beyond the Sequence: Statistics-Driven Pre-training for Stabilizing Sequential Recommendation Model}
\author{Sirui Wang}
\authornote{Both authors contributed equally to this research.}
\authornote{Corresponding author.}
\affiliation{
  \institution{Department of Automation, Tsinghua University}
  \city{Beijing}
  \country{China}
}
\affiliation{
  \institution{Meituan}
  \city{Beijing}
  \country{China}
}
\email{wangsirui@meituan.com}

\author{Peiguang Li}
\authornotemark[1]
\affiliation{
  \institution{Meituan}
  \city{Beijing}
  \country{China}
}
\email{lipeiguang@meituan.com}

\author{Yunsen Xian}
\affiliation{
  \institution{Meituan}
  \city{Beijing}
  \country{China}
}
\email{xianyunsen@meituan.com}

\author{Hongzhi Zhang }
\affiliation{
  \institution{Meituan}
  \city{Beijing}
  \country{China}
}
\email{zhhongzhi2015@gmail.com}

\renewcommand{\shortauthors}{Wang and Li et al.}

\begin{abstract}
The sequential recommendation task aims to predict the item that user is interested in according to his/her historical action sequence. However, inevitable random action, i.e. user randomly accesses an item among multiple candidates or clicks several items at random order, cause the sequence fails to provide stable and high-quality signals. To alleviate the issue, we propose the \underline{S}tatis\underline{T}ics-\underline{D}riven \underline{P}re-traing framework (called STDP briefly). The main idea of the work lies in the exploration of utilizing the statistics information along with the pre-training paradigm to stabilize the optimization of recommendation model. Specifically, we derive two types of statistical information: item co-occurrence across sequence and attribute frequency within the sequence. And we design the following pre-training tasks: 1) The co-occurred items prediction task, which encourages the model to distribute its attention on multiple suitable targets instead of just focusing on the next item that may be unstable. 2) We generate a paired sequence by replacing items with their co-occurred items and enforce its representation close with the original one, thus enhancing the model’s robustness to the random noise. 3) To reduce the impact of random on user’s long-term preferences, we encourage the model to capture sequence-level frequent attributes. The significant improvement over six datasets demonstrates the effectiveness and superiority of the proposal, and further analysis verified the generalization of the STDP framework on other models.
\end{abstract}




\maketitle

\section{Introduction}
Online platforms play an indispensable part of current daily lives with providing us various and ample items, but still struggle with recommending interested ones for customer. To alleviate this problem, the sequential recommendation task is proposed to distill users' interests from their historical action sequences thus conduct appropriate recommendation~\cite{hidasi2015session, SASRec, hao2023feature}. Typically, the action sequence is composed of items that user accessed (click, buy, etc.) and both the long-term preferences and near-term dynamic favor embodied in the sequence impact the user's interests of the next item~\cite{lv2019sdm, zheng2022disentangling}, which make the sequential recommendation a challenging task.

Existing works exploit various approaches to capture user interests from action sequences. Gated Recurrent Units (GRU) is first introduced~\cite{GRU4Rec} and followed by a series of improved versions~\cite{knowledge4rec, GRU4Rec_Hierarchical}. There are also methods that attempt other networks such as Graph Neural Network (GNN)~\cite{yu2022graph, wu2022graph} and Transformer~\cite{SASRec, bert4rec, ma2023improving}, as well as training paradigms including Contrastive Learning~\cite{cl_rec, liu2021contrastive} and Self-Supervision Learning~\cite{Zhou_S3, tian2022temporal, lin2023self}. Despite the promising progress achieved with previous methods, there still a major challenge stems from the random noise remain to be solved. As observed in real-world recommendation platforms, random noise might come from users randomly select an item from multiple suitable candidates, users access several items at random orders, etc. Such random noise cause the sequences cannot provide stable supervision signals for describing user preferences and further disrupts the optimization of the model. 

How to deal with noise issue attract much attention from the research community. \citeauthor{unreliable_instance} first verify the existence and severity of the noisy action and propose an filter to remove unreliable instances. \citeauthor{zhang2021causerec} propose to model the counterfactual data distribution to confront the sparsity and noise nature of action sequence. And, \citeauthor{lin2023self} propose a two-stage method, they first corrects
the sequence by adjusting the noise items then train the model with the corrected sequence. However, we argue that the existing methods relies on locating noise item/action and bringing error accumulation problem, and random items also contains valuable information describing user preferences.

Since random noise is inevitable, in this work, we start from reducing their impact on model optimization with more stable information. Specifically, we notice that although an individual item may contain random and unstable features, the statistical group it belongs to has more stable information. Further, we propose the \underline{S}tatis\underline{T}ics-\underline{D}riven \underline{P}re-traing (STDP) framework, which utilizes the statistics information along with the pre-training paradigm to stabilize the optimization of recommendation model. It is worth noting that the stabilizing here is considered as: we expect to use statistical information as an additional supervision signal to make the optimization process robust to noise and more stable. From this perspective, it can be considered that our method still belongs to the field of denoising.

First of all, we derive the item co-occurrence across sequence via statistics from the training data (only the training data is counted to prevent data leakage) and devise two pre-training tasks: 1) Co-occurred Items Prediction (CIP). We optimize the model to predict the next item along with its top co-occurred items, thus encourage the model to allocate its attention on multiple suitable targets instead of just focusing on the next item that may be unstable. 2) Paired Sequence Similarity (PSS). We randomly replace part of items in the original sequence with their co-occurred items and obtain the paired sequence then maximize the similarity between the representation of original and paired sequences. In this manner, the model's robustness is enhanced via imitating the random access in the inputs. Moreover, we collect the attribute frequency information within the sequence via statistics. On the basis of the Item associated Attribute Prediction task is proven to be helpful~\cite{Zhou_S3}, we further design an sequence-level Frequent Attribute Prediction (FAP) task to encourage the model predicting the frequent attributes based on sequential features, facilitating the capturing of stable user's long-term preferences.

The major contributions of this paper are summarized as follows: (1) We reveal the impact of random noise and argue they cause sequences cannot provide stable supervision signals. To alleviate this issue, we present the \underline{S}tatis\underline{T}ics-\underline{D}riven \underline{P}re-traing (STDP)  framework to promote the recommendation models via pre-training on more stable statistics information. (2) In the STDP framework, several pre-training tasks are designed with statistics information to reduce the effect of random noise in different aspects: distribute models' focus on multiple suitable targets, improve the robustness of sequence representation, and capture stable long-term preferences. (3) Experimental results verify the effectiveness of our proposed STDP framework, which improves the existing method to achieve state-of-the-art performance. The extensive analysis reveals the generalization of the proposal over other recommendation models.

\section{Related Work}

General recommendation systems mine the implicit association between users and items from users' historical interactions. In the line of works based on the Collaborative Filtering~\cite{Item_CF}, Matrix Factorization~\cite{MF} is the most representative method, which represent user and item with unified embeddings and calculate their similarity to estimate user's selection. In recent years, the benefits of various auxiliary information are widely exploited, including textual~\cite{text_CF} and visual~\cite{image_CF} etc. 

Different from the general recommendation, the sequential recommendation task requires modeling the users’ historical behaviors in order to predict next item. The pioneer works~\cite{fuse_mc} utilize the Markov Chain (MC) assumption to estimate the item-item transition matrix. In the line of works that adopt various neural networks like GRU~\cite{GRU4Rec, knowledge4rec, GRU4Rec_Hierarchical}, Convolutional Neural Network (CNN)~\cite{Caser}, GNN~\cite{yu2022graph, wu2022graph}, and Transformer~\cite{SASRec, bert4rec, ma2023improving}, achieve promising results. Recently, the works~\cite{FDSA,Zhou_S3,cross_rec} leverage auxiliary information and employ Contrastive Learning~\cite{cl_rec, liu2021contrastive} and Self-Supervision Learning~\cite{Zhou_S3, tian2022temporal, lin2023self} training paradigms achieve advanced performance, which also inspired our work.

There are also several methods that relevant with our work.
\citeauthor{xie2023multi} design several multi-granularity contrastive learning tasks, among which the session-level task also utilize the global co-occurrences of items that directly counted from all sessions. Our work differs from this approach  in the following aspects: 1) We use Jaccard distance~\ref{equ6} to measure the degree of correlation between items to avoid the interference of high-frequency items, because the directly counted times of co-occurrences between items is easily disturbed by high-frequency repurchase and unrelated items, such as toilet paper and toothpaste. 2) Our statistics are directional, which means we only count successors of the item to avoid the problem of cyclic repetition. For example, suppose that a camera has been already purchased in previous and the next item is a lens, if we still require the model to output/predict a camera may confuse the model. \citeauthor{cl_rec} propose some augmentation operations including item crop, mask, and reorder. On the basis of item reorder, our PSS task further uses the co-occurred item to replace the original item to construct paird sequence.

\section{Proposed Method}

\subsection{Approach}

\subsubsection{Overview.} 

\begin{figure}[h]
	\centering 
        \includegraphics[width=1.0\linewidth]{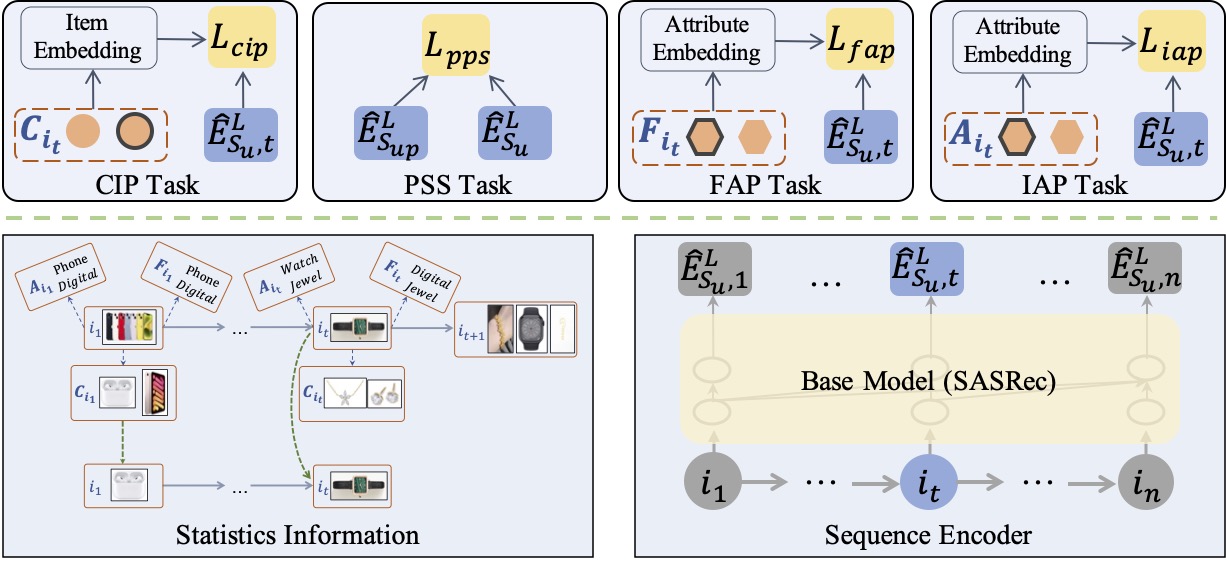}
	\caption{The structure of \underline{S}tatis\underline{T}ics-\underline{D}riven \underline{P}re-traing (STDP) framework. Bottom panel illustrates examples of statistics and how a paired sequence is created, then a base model is employed to model the original sequence and the paired sequence separately. The top panel illustrates pre-training tasks: Co-occurred Items Prediction (CIP), Paired Sequence Similarity (PSS), Frequent Attribute Prediction (FAP), and Item Attribute Prediction (IAP).}
	\label{method}
\end{figure}

The purpose of our approach is utilizing relatively stable statistics information to reduce the impact of random noise. To this end, we derive the co-occurred items and frequent attributes via statistics and propose STDP framework. Overview the STDP framework is shown in Figure~\ref{method}. As we can see, the framework adopts SASRec as backbone model and accompanied with several pre-training tasks: Co-occured Item Prediction (CIP), Paired Sequence Similarity (PSS), Frequent Attribute Prediction (FAP), and Item Attribute Prediction (IAP) proposed by \citeauthor{Zhou_S3}. In the following subsections, we first briefly review the base method, then we describe the details about pre-training tasks.

\subsubsection{Base Method.} The SASRec model mainly consists of an Embedding layer, Encoding layer, and Prediction layer. Due to space limitations, only a brief introduction is given here, more details can be referred to~\cite{SASRec} and ~\cite{Zhou_S3} et al.

Given the sets of users $\mathcal{U}$, items $\mathcal{I}$, and items' attributes $\mathcal{A}$, the sequence $S_{u} = \{i_{1}, i_{2}, \dots, i_{n} \}$ is composed of the items that user $u \in \mathcal{U}$ has accessed and organized in the chronologically order, each item $i \in \mathcal{I}$ is associated with several attributes $A_{i} = \{ a_{1}, a_{2}, \dots, a_{m} \}$. In Embedding layer, the sequence $S_{u}$ with $n$ items is mapped into a $d$-dimensional vector sequence $\mathbf{E}_{S_{u}} \in R^{n \times d}$, and then the position features is integrated with $\mathbf{E}_{S_{u}}$ to supplement the position information.

Encoding layer employs the stacked $L$ self-attention blocks to perform sequential encoding for $\mathbf{E}_{S_{u}}$ and produce $\hat{\mathbf{E}}_{S_{u}}^{L}$. Each self-attention block is composed of a multi-head self-attention module and a point-wise feed-forward module. The multi-head self-attention module aims to capture the correlation among items from diverse $H$ sub-spaces, and the feed-forward module supplements the non-linearity to features in the self-attention results.

In the prediction layer, the probability that the user $u$ to access item $i$ in the next step $t+1$ is calculated as:
\begin{equation}
\label{equ5}
P(i|{S_{u}}_{1:t}) = i_{e} \cdot \hat{\mathbf{E}}_{S_{u},t}^{L} \,
\end{equation}

\subsubsection{Statistics-driven pre-training tasks.}
For each item $i$, we follow the Jaccard criterion to extract its top-$k$ co-occurred items $C_{i}$ across all sequences. To ensure co-occurred items are complementary for each other and prevent item from appearing recursively, the co-occurred items are selected from $i$'s successor items in the sequence. The Jaccard score between item $i_{a}$ and $i_{b}$ is formulated as:
\begin{equation}
\label{equ6}
Jaccard(i_{a}, i_{b}) = \frac{cnt(i_{a}, i_{b})}{cnt(i_{a}) + cnt(i_{b}) - cnt(i_{a}, i_{b})} \,
\end{equation}
where $cnt(i_{a})$, $cnt(i_{b})$, and $cnt(i_{a}, i_{b})$ indicate the number of sequences in the dataset that containing $i_{a}$, $i_{b}$, and $i_{b}$ belong to successor items of $i_{a}$, respectively. Notably, to avoid data leakage, our statistics are performed on the first $n-2$ item segment of each sequence.

As interpreted ahead, the random action, e.g. user select an item randomly from multiple candidates, cause the sequences cannot provide high-quality supervision signals. Its reflection on the training process is: the biased supervision signal makes the model training unstable. Therefore, we argue that external forces are needed to promote the model via distributing its focus on the multiple suitable candidates instead of the picked single one. Specifically, we propose the \textbf{Co-occurred Item Prediction} (CIP) task. Given the item $i$ at step $t$ and its co-occurred items set $C_{i}$, we randomly select a positive target $i_{p}$ from $C_{i}$ and a negative target $i_{n}$ from the difference set $\mathcal{I}-C_{i}$ at each step of pre-training. Then we adopt the pairwise rank loss as the objective function to joint optimize the model on this task:
\begin{equation}
\label{equ7}
L_{cip} = -\sum_{t} log( \sigma (P(i_{p}|{S_{u,1:t}})) - \sigma (P(i_{n}|{S_{u,1:t}}))) \,
\end{equation}

Furthermore, since the ultimate goal of model is to learn sequence representation, we present another item-oriented pre-training task called the \textbf{Paired Sequence Similarity} (PSS). Concretely, each item in the original sequence $S_{u}$ is replaced randomly with one of its co-occurred items with the probability $p_{rpc}$. After replacing, we exchange the order of items randomly to simulate the situation that the user access several items in random order. In this manner, we obtain a paired sequence $S_{up}$ and its' representation $\hat{\mathbf{E}}_{S_{up}}^{L}$ from the same SASRec.

To strengthen the model's robustness to random noise, referring to the R-drop~\cite{rdrop}, we employ the bidirectional Kullback-Leibler (KL) divergence to measure the similarity between two sequences:
\begin{equation}
\label{equ8}
L_{pss} = \sum_{t} \frac{1}{2}(KL(\hat{\mathbf{E}}_{S_{u, 1:t}}^{L}||\hat{\mathbf{E}}_{S_{up, 1:t}}^{L}) + KL(\hat{\mathbf{E}}_{S_{up, 1:t}}^{L}||\hat{\mathbf{E}}_{S_{u, 1:t}}^{L})) \,
\end{equation}
where $KL$ is calculated at each dimensional $j$ of the features:
\begin{equation}
\label{equ9}
KL(e_{1}||e_{2}) = -\sum_{j}^{d}e_{1,j}(log(e_{2,j} - e_{1,j})) \,
\end{equation}

Finally, to explicitly facilitate the model capturing the user's long-term preference, we design a sequence-level task named \textbf{Frequent Attribute Prediction} (FAP). Before that, to inject the attribute information into item representation, we borrow the \textbf{Item Attribute Prediction} (IAP) task proposed by \citeauthor{Zhou_S3}. In IAP task, given the item $i$ at step $t$ and its attribute set $A_{i}$, a positive target $a_{p}$ and a negative target $a_{n}$ is selected randomly from $A_{i}$ and difference set $\mathcal{A}-A_{i}$, separately, then a pairwise rank loss is adopted to optimize model to distinct $a_{p}$ and $a_{n}$:
\begin{equation}
\label{equ10}
L_{iap} = -\sum_{t} log( \sigma (P(a_{p}|S_{u,1:t})) - \sigma (P(a_{n}|S_{u,1:t}))) \,
\end{equation}

In FAP task, at step $t$, we intercept the first $t$ items in the sequence $S_{u}$ as the segment $S_{u, 1:t}$, and we select the top-$k$ attributes that appear most frequently in the attribute sets of all items as the long-term-preferred attribute set $F_{t}$ of the segment. Likewise, for the segment $S_{u, 1:t}$ and its preferred attributes set $F_{t}$, a positive target $a_{p}$ and a negative target $a_{n}$ is selected randomly from $F_{t}$ and difference set $\mathcal{A}-F_{t}$, respectively, and the pairwise rank loss is set as the objective function:
\begin{equation}
\label{equ11}
L_{fap} = -\sum_{t} log( \sigma (P(a_{p}|S_{u, 1:t})) - \sigma (P(a_{n}|S_{u, 1:t})) )\,
\end{equation}

\subsubsection{Optimization.} 
During the pre-training, we combine four tasks described above in the joint learning way, and the final objective function is formulated as:
\begin{equation}
\label{equ13}
L_{pre} = \lambda_{cip} \cdot L_{cip} + \lambda_{pss} \cdot L_{pss} + \lambda_{iap} \cdot L_{iap} + \lambda_{fap} \cdot L_{fap} \,
\end{equation}
where $\lambda_{*}$ denote the weights of auxiliary tasks.

After pre-training, \textbf{Next Item Prediction} (NIP) task is the only the objective in fine-tuning stage:
\begin{equation}
\label{equ12}
L_{nip} = -\sum_{t} log( \sigma ( P(i_{p}|{S_{u,1:t}})) - \sigma (P(i_{n}|{S_{u,1:t}}))) \,
\end{equation}
where $i_{p}$ and $i_{n}$ indicates the ground-truth item and negative item. 

\section{Experiments}

\subsection{Experimental Settings}
\begin{table}
    \centering
    \begin{center}
        \caption{Statistics of the experimental datasets.}
        \label{Dataset_statistic_info} 
        \begin{tabular}{l|cccc}
            \hline
            \hline
            Datasets & \#Seq & \#Items  & \#Items/Seq & \#Attrs/Item \\
            \hline
            Meituan & 13,622 & 20,062 & 54.9 & 8.8  \\
            Beauty & 22,363 & 12,101 & 8.9 & 3.7  \\
            Sports & 25,598 & 18,357 & 8.3 & 6.0  \\
            Toys & 19,412 & 11,924 & 8.6 & 4.3  \\
            Yelp & 30,431 & 20,033 & 10.4 & 4.8  \\
            LastFM & 1,090 & 3,646 & 48.2 & 31.5  \\
            \hline
            \hline
        \end{tabular}
    \end{center}
\end{table}

\begin{table*}
	\centering
	\begin{center}
        \caption{Performance comparison on six datasets, the best values are highlighted in \textbf{bold} and the second-best results are marked with \underline{underline}. Ipv$_{best}$ and Ipv$_{base}$ indicates the improvements over the best published baseline and base method (SASRec$_{IAP}$), respectively, and all improvements over the best baseline are statistically significant (t-tests, p < 0.05).}
	\label{main_results} 
		\begin{tabular}{l | l | c c c c c c c c}
			\hline
			\hline
			Datasets & Metric & SASRec & BERT4Rec & SASRec$_{IAP}$ & FDSA & S$^{3}$-Rec & Proposed & Ipv$_{best}$ & Ipv$_{base}$ \\
			\hline
            \multirow{5}{*}{Meituan}&
			HR@5 & 0.4524 & 0.3985 & 0.4782 & 0.4595 & \underline{0.4925} & \textbf{0.5250} & 6.60\% & 9.79\% \\
			~ & NDCG@5 & 0.3207 & 0.2713 & 0.3473 & 0.3236 & \underline{0.3527} & \textbf{0.3911} & 10.89\% & 12.61\% \\
			~ & HR@10 & 0.6053 & 0.5514 & 0.6190 & 0.6164 & \underline{0.6368} & \textbf{0.6623} & 4.00\% & 7.00\% \\
			~ & NDCG@10 & 0.3700 & 0.3208 & 0.3927 & 0.3743 & \underline{0.3994} & \textbf{0.4355} & 9.04\% & 10.90\% \\
			~ & MRR & 0.3146 & 0.2689 & 0.3394 & 0.3167 & \underline{0.3421} & \textbf{0.3798} & 11.02\% & 11.90\% \\
			\hline
			\multirow{5}{*}{Beauty}&
			HR@5 & 0.3741 & 0.3640 & 0.4126 & 0.4010 & \underline{0.4502} & \textbf{0.4677} & 3.89\% & 13.35\% \\
			~ & NDCG@5 & 0.2848 & 0.2622 & 0.3183 & 0.2974 & \underline{0.3407} & \textbf{0.3603} & 5.75\% & 13.20\% \\
			~ & HR@10 & 0.4696 & 0.4739 & 0.5106 & 0.5096 & \underline{0.5506} & \textbf{0.5687} & 3.29\% & 11.38\% \\
			~ & NDCG@10 & 0.3156 & 0.2975 & 0.3499 & 0.3324 & \underline{0.3732} & \textbf{0.3930} & 5.31\% & 12.32\% \\
			~ & MRR & 0.2852 & 0.2614 & 0.3165 & 0.2943 & \underline{0.3340} & \textbf{0.3535} & 5.84\% & 11.69\% \\
			\hline
            \multirow{5}{*}{Sports}&
			HR@5 & 0.3466 & 0.3375 & 0.3769 & 0.3855 & \underline{0.4267} & \textbf{0.4443} & 4.03\% & 17.88\% \\
			~ & NDCG@5 & 0.2497 & 0.2341 & 0.2774 & 0.2756 & \underline{0.3104} & \textbf{0.3293} & 6.09\% & 18.71\% \\
			~ & HR@10 & 0.4622 & 0.4722 & 0.4952 & 0.5136 & \underline{0.5614} & \textbf{0.5678} & 1.14\% & 14.66\% \\
			~ & NDCG@10 & 0.2869 & 0.2775 & 0.3155 & 0.3170 & \underline{0.3538} & \textbf{0.3693} & 4.38\% & 17.05\% \\
			~ & MRR & 0.2520 & 0.2378 & 0.2785 & 0.2748 & \underline{0.3071} & \textbf{0.3247} & 5.73\% & 16.59\% \\
			\hline
			\multirow{5}{*}{Toys}&
			HR@5 & 0.3682 & 0.3344 & 0.4039 & 0.3994 & \underline{0.4420} & \textbf{0.4620} & 4.52\% & 14.38\% \\
			~ & NDCG@5 & 0.2820 & 0.2327 & 0.3117 & 0.2903 & \underline{0.3270} & \textbf{0.3555} & 8.72\% & 14.05\% \\
			~ & HR@10 & 0.4663 & 0.4493 & 0.5049 & 0.5129 & \underline{0.5530} & \textbf{0.5632} & 1.84\% & 11.55\% \\
			~ & NDCG@10 & 0.3136 & 0.2698 & 0.3442 & 0.3271 & \underline{0.3629} & \textbf{0.3882} & 6.97\% & 12.78\% \\
			~ & MRR & 0.2842 & 0.2338 & 0.3114 & 0.2863 & \underline{0.3202} & \textbf{0.3495} & 9.15\% & 12.24\% \\
			\hline  
            \multirow{5}{*}{Yelp}&
			HR@5 & 0.5745 & 0.5976 & 0.5935 & 0.5728 & \underline{0.6085} & \textbf{0.6508} & 6.95\% & 9.65\% \\
			~ & NDCG@5 & 0.4113 & 0.4252 & 0.4312 & 0.4014 & \underline{0.4401} & \textbf{0.4881} & 10.91\% & 13.20\% \\
			~ & HR@10 & 0.7373 & 0.7597 & 0.7447 & 0.7555 & \underline{0.7725} & \textbf{0.7952} & 2.94\% & 6.78\% \\
			~ & NDCG@10 & 0.4642 & 0.4778 & 0.4803 & 0.4607 & \underline{0.4934} & \textbf{0.5351} & 8.45\% & 11.41\% \\
			~ & MRR & 0.3927 & 0.4026 & 0.4107 & 0.3834 & \underline{0.4190} & \textbf{0.4644} & 10.84\% & 13.08\% \\
			\hline
			\multirow{5}{*}{LastFM}&
			HR@5 & 0.3385 & 0.3569 & 0.3404 & 0.2624 & \underline{0.4523} & \textbf{0.4606} & 1.84\% & 35.31\% \\
			~ & NDCG@5 & 0.2330 & 0.2409 & 0.2279 & 0.1766 & \underline{0.3156} & \textbf{0.3296} & 4.44\% & 44.62\% \\
			~ & HR@10 & 0.4706 & 0.4991 & 0.4817 & 0.4055 & \underline{0.5835} & \textbf{0.6055} & 3.77\% & 25.70\% \\
			~ & NDCG@10 & 0.2755 & 0.2871 & 0.2736 & 0.2225 & \underline{0.3583} & \textbf{0.3766} & 5.11\% & 37.65\% \\
			~ & MRR & 0.2364 & 0.2424 & 0.2296 & 0.1884 & \underline{0.3072} & \textbf{0.3225} & 4.98\% & 40.46\% \\
			\hline
			\hline
		\end{tabular}
	\end{center}
\end{table*}

\subsubsection{Datasets.} Six public datasets collected from Meituan\footnote{https://www.meituan.com}, Amazon (including Beauty, Sports, and Toys), Yelp\footnote{https://www.yelp.com/dataset}, and LastFM\footnote{https://grouplens.org/datasets/hetrec-2011/} platforms, respectively, are selected as the benchmarks for evaluation. Apart from the Meituan dataset that provides categories, locations, and keywords as the attributes of the items, the rest datasets only remain categories as the attributes. The experimental data is published by S$^{3}$-Rec~\cite{Zhou_S3}, in which the item sequences are organized in chronological order. The detailed statistics of these datasets are listed in Table~\ref{Dataset_statistic_info}.

\subsubsection{Implementation and Evaluation.} For the data preparation, the size of co-occurred items set, attribute set, and favorite set are all limited to 20, the replace rate in PSS task is set as 0.2. The sequence lengths are padded to 50 and the size of each mini-batch is padded to 256. During the training, the hyper-parameters $\lambda$ is choose from the range \{0.1, 0.3, 0.5, 0.8, 1.0\} independently and finally $\lambda_{1}, \lambda_{2}, \lambda_{3}, \lambda_{4}$ are set to 0.3, 0.3, 0.8, 0.5, respectively.

To make a fair comparison with the previous methods~\cite{FDSA,Zhou_S3}, the last and the second-to-last item in each sequence is reserved for test and validation, and the remaining items are used for training. We adopt Hit Ratio cutoff at 5 and 10 (HR@5, HR@10), Normalized Discounted Cumulative Gain cutoff at 5 and 10 (NDCG@5, NDCG@10), and Mean Reciprocal Rank (MRR) as metrics. Notably, the evaluation is based on the sampled item set, which is obtained by pairing the ground-truth item with 99 negative items that randomly sampled from the other sequences.

\subsubsection{Baseline Methods.} We select the following competitive baseline methods for performance comparison: \textbf{SASRec}~\cite{SASRec} transfers multi-head self-attention mechanism from Transformer network to capture long-term semantics, and its improved versions $\textbf{SASRec}_{IAP}$ which pre-trained with IAP task. \textbf{BERT4Rec}~\cite{bert4rec} adopts bidirectional self-attention mechanism and pre-trained with Cloze objective loss. \textbf{FDSA}~\cite{FDSA} integrates heterogeneous information into item sequences and predicts the next item from multiple perspectives. \textbf{S}$^{3}$-\textbf{Rec}~\cite{Zhou_S3} pre-train the SASRec model to capture the correlation between the heterogeneous information for improving sequential recommendation, achieving state-of-the-art performance under current evaluation system.

\subsection{Experimental Results}

\subsubsection{Overall Performance.}
Table~\ref{main_results} reports the performance of various methods on the target datasets. Overall, on MRR metric, our proposal promote the base method SASRec$_{IAP}$ significantly with an average $17.66\%$ improvement, which also outperforms the previous best method with an average $7.93\%$ advantage, presenting a solid state-of-the-art performance. Analyzing the results in Table~\ref{main_results}, we can infer:

Firstly, it's notable that SASRec$_{IAP}$ outperforms FDSA method, which takes attributes as input directly. We believe that items and attributes express the distinct aspect of information in the sequential recommendation task, and the inconsistency between them restrains the attribute information to bring a stable improvement. Further, S$^{3}$-Rec surpasses the existing methods and achieves previous stat-of-the-art performance. The superiority of S$^{3}$-Rec mainly comes from it 1) introduces the intrinsic data correlation, including attributes level and item level, and 2) incorporates the pre-training techniques to fuse various information to enhance item representations, which also inspire our work. 

Secondly, our proposal outperforms baseline methods and achieves the best scores on all datasets. Based on the above analyses, we summarize the advantages of our method to the following aspects: 1) Our work is the first to attempt to utilize the statistics information to reduce the negative impact of random noise, especially we encourage the model to distribute its focus on multiple targets and improve model's robustness to the randomness at sequence-level. 2) Rather than utilizing the attributes knowledge as in previous methods~\cite{FDSA,Zhou_S3}, we investigate the contribution of frequent attributes that appear in the sequence, which facilitate the model to capture user’s long-term preferences.

\subsubsection{Ablation Study.}
To probe the influence of the proposed tasks in STDP, we design a series of variants and report their HR@10, NDCG@10, and MRR scores on Meituan, Beauty, and Yelp datasets in Table~\ref{ablation_results}.

In fact, the work adopts a multi-task paradigm in the early stage and the results are shown in first group in Table~\ref{ablation_results} (- Pre-train). It can be found that the results of the multi-task are lower than pre-training on most scores, and we guess this is caused by that multi-task disperses the optimization goals. Next, the second group reports the results of removing item-related tasks. As we can infer, removing PSS task(-PSS) appears to have a larger impact on performance than removing CIP task(-CIP). Given that the PSS task acts on the sequence level, thus its contributes to the sequential representation is more directly than CIP task that works at the item level. Also, removing both CIP and PSS(-CIP$\&$PSS) decreases the performance significantly to the lowest among all variants, validating the essence of global co-occurrence information and corresponding strategies.

\begin{table*}
    \centering
    \begin{center}
        \caption{Ablation results on Meituan, Beauty, and Yelp datasets. Our final results are written in italics. ``-'' indicates removing the selected strategy while reserving the rest.}
        \label{ablation_results} 
        \begin{tabular}{lccccccccc}
            \hline
            \hline
            \multirow{2}{*}{Methods}&
            \multicolumn{3}{c}{Meituan}&
            \multicolumn{3}{c}{Beauty} &
            \multicolumn{3}{c}{Yelp} \\
            \cmidrule[0.05em](lr){2-4} \cmidrule[0.05em](lr){5-7} \cmidrule[0.05em](lr){8-10}
            & HR@10 & NDCG@10 & MRR
            & HR@10 & NDCG@10 & MRR
            & HR@10 & NDCG@10 & MRR \\
            \hline
            Proposed & \textit{0.6623} & \textit{0.4355} & \textit{0.3798} & \textit{0.5687} & \textit{0.3930} & \textit{0.3535} & \textit{0.7952} & \textit{0.5351} & \textit{0.4644} \\
            \hline
            - Pre-train & 0.6603 & 0.4273 & 0.3701 & 0.5567 & 0.3901 & 0.3536 & 0.7998 & 0.5325 & 0.4597 \\
            \hline
            - CIP & 0.6590 & 0.4261 & 0.3688 & 0.5542 & 0.3849 & 0.3476 & 0.7918 & 0.5322 & 0.4618 \\
            - PSS & 0.6366 & 0.4068 & 0.3519 & 0.5379 & 0.3714 & 0.3357 & 0.7690 & 0.5050 & 0.4343 \\
            - (CIP\&PSS) & 0.6250 & 0.3962 & 0.3417 & 0.5329 & 0.3680 & 0.3329 & 0.7621 & 0.4972 & 0.4268 \\
            \hline
            - FAP & 0.6608 & 0.4307 & 0.3742 & 0.5510 & 0.3853 & 0.3493 & 0.7950 & 0.5279 & 0.4556 \\
            - IAP & 0.6627 & 0.4290 & 0.3713 & 0.5510 & 0.3824 & 0.3456 & 0.7960 & 0.5344 & 0.4635 \\
            - (IAP\&FAP)  & 0.6567 & 0.4227 & 0.3653 & 0.5305 & 0.3707 & 0.3370 & 0.7924 & 0.5249 & 0.4524 \\
            \hline
            \hline
        \end{tabular}
    \end{center}
\end{table*}

Besides, we explore the effect of the attributes-related tasks and list the results in the second group of Table~\ref{ablation_results}. Firstly, removing FAP(-FAP) brings a slight performance degradation. Given that both IAP and FAP utilize attribute information and FAP is considered the sequence-level version of IAP. Hence we doubt that current experimental results are influenced by the IAP task. To verify this point, we compare the results of removing IAP (-IAP) and removing both of IAP and FAP (-IAP\&FAP) and draw the following conclusion: 1) Results margin between -IAP and -IAP\&FAP reveals the usefulness of proposed FAP task. 2) The contribution of attribute information is confirmed by the results that removing both IAP and FAP causes significant performance degradation.

\subsubsection{Generalization Verification.}
\begin{figure*}
    \centering
    \caption{Performance comparison between GRU4Rec (blue), GRU4Rec with IAP (green), and GRU4Rec with STDP framework (red).}
    \label{generalization}
    \subfigure[Meituan]{
        \centering
        \includegraphics[width=4.8cm]{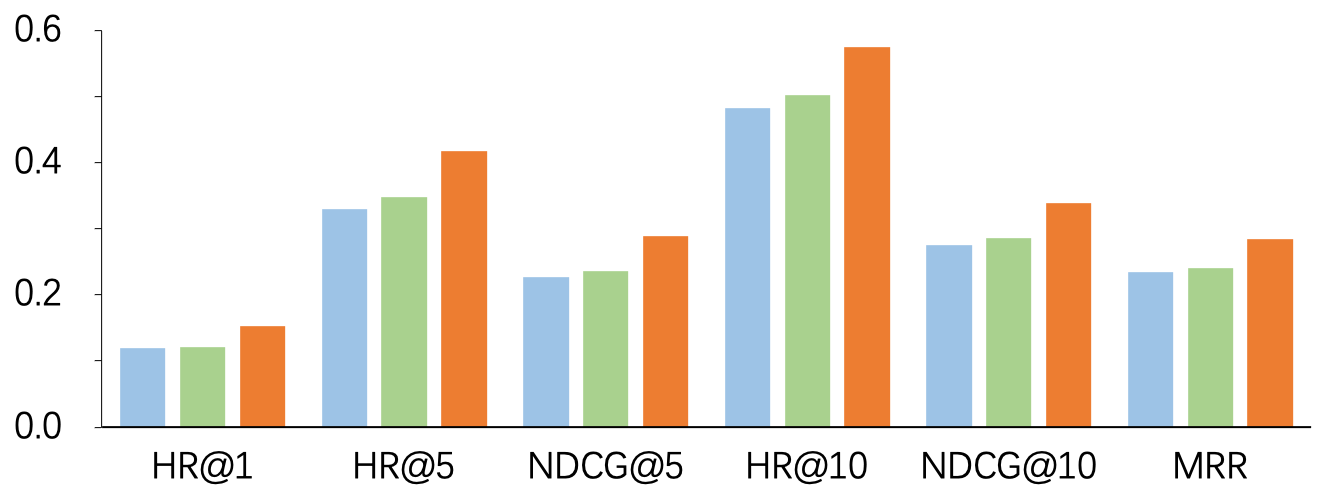}
    }
    \subfigure[Beauty]{
        \centering
        \includegraphics[width=4.8cm]{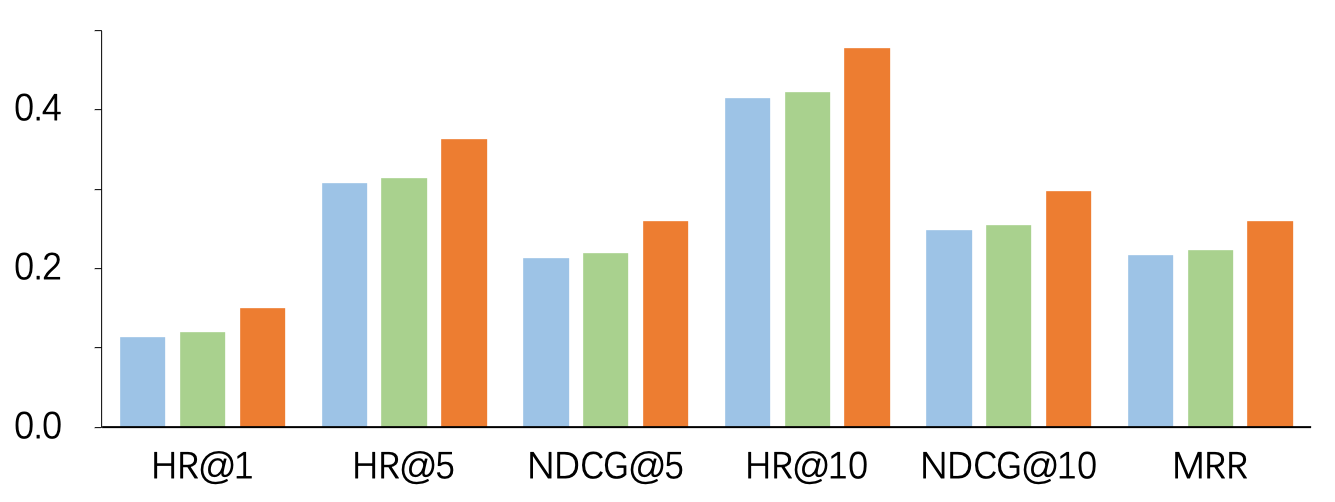}
    }
    \subfigure[Yelp]{
        \centering
        \includegraphics[width=4.8cm]{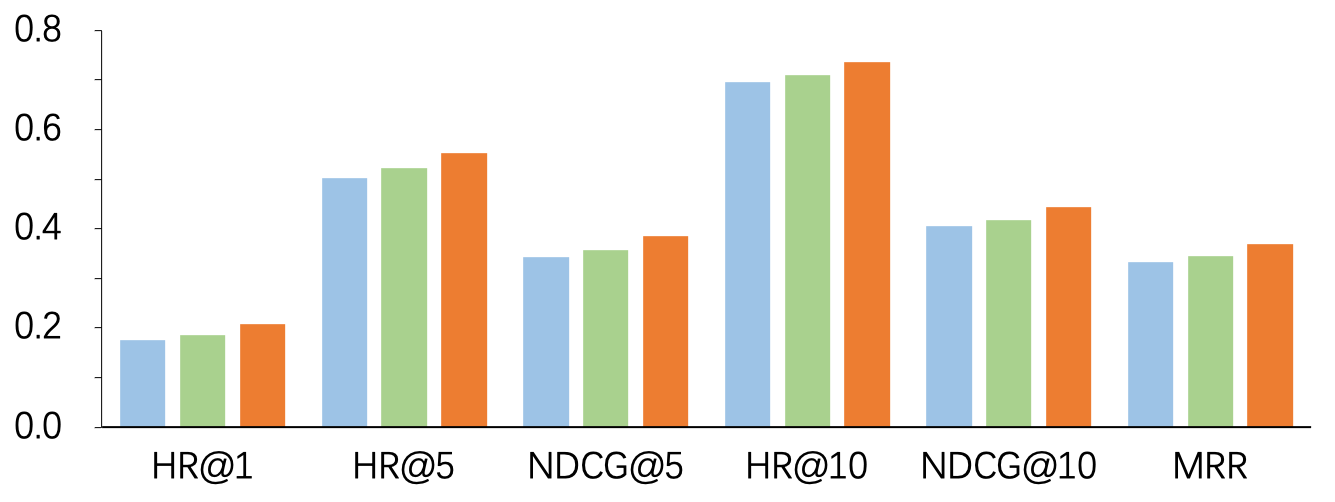}
    }
\end{figure*}

As we introduced before, our proposed STDP framework are not customized for any specific sequential recommendation model. To verify this, we add our proposed strategies to the GRU4Rec~\cite{GRU4Rec} model to examine the performance influence. Referring to the codes public available~\footnote{https://github.com/hungthanhpham94/GRU4REC-pytorch}, we re-implementation the GRU4Rec method and pre-train it using the STDP directly without further tuning. As illustrated in Figure~\ref{generalization}, the performances of GRU4Rec are improved obviously by combining STDP, demonstrating the effectiveness and generalization of our designed STDP framework.

\section{Conclusion}
In this paper, we first exploit the impact of random noise within the action sequence and we argue such randomness causes the sequence cannot provide stable and high-quality supervision signals for recommendation model. To alleviate this issue, we propose a simple yet effective framework, which utilizes stable statistics information and pre-training tasks to reduce the influence of randomness. The proposal promotes the base method with an average improvement of 17.66\%, surpassing the previous best result with 7.93\% on six datasets. Also, the consistent improvements on other model demonstrate the generalization of the proposal. Due to the limitations of datasets, codes, evaluation system, etc., our work has not been compared with other related methods, but we believe the current results are sufficient to illustrate its effectiveness. We would continue current work to perform further research and hope our work will draw the attention of the community to the issue of random noise.

\bibliographystyle{ACM-Reference-Format}
\bibliography{sample-base}

\end{document}